\documentclass[proceedings]{JHEP3}

\PrHEP{PrHEP hep2001}                   
\conference{International Europhysics Conference on HEP}
\usepackage{epsfig}                   
\def\gev{\ifmmode \mathop{\rm GeV}\nolimits\else {\rm GeV}\fi}
\def\mev{\ifmmode \mathop{\rm MeV}\nolimits\else {\rm MeV}\fi}

\title{Precision Determination
of the Muon $g_{\mu}-2$ and $\bar{\alpha}_{\rm Q.E.D.}(M^2_{Z})$
\footnote{To appear in the Proceedings of the Budapest 2001 EPS HEP 
Conference}}

\author{J. F. de Troc\'oniz and \speaker{F. J. Yndur\'ain}\\
        Dept. de F\'{\i}sica Te\'orica, C-XI,\\
        Universidad Aut\'onoma de Madrid,\\
        E-28049, Cantoblanco (Madrid), Spain.\\
        E-mail: \email{fjy@delta.ft.uam.es}}

\abstract{We present a summary of the results of two recent precise
calculations of the muon anomalous magnetic moment and the electromagnetic
coupling on the $Z$.}

\begin{document}

\section{Introduction}
We present a summary of the results of two recent precise
calculations of the muon anomalous magnetic moment ($g_{\mu}-2$)~\cite{g2} and 
the electromagnetic coupling on the $Z$ ($\bar{\alpha}_{\rm Q.E.D.}(M^2_{Z})$)~\cite{az}.
The main sources of uncertainty are underlined.

\section{Calculation of the muon $g_{\mu}-2$}
The appearance of a new, very precise measurement of the muon magnetic
moment~\cite{uno} has triggered the interest in theoretical calculations 
of this quantity. 
Particularly, because the experimental figure (we give the result 
for the anomaly, averaged with older determinations~\cite{dos})
\begin{equation}
a_\mu(\hbox{exp.})= \frac{g_{\mu}-2}{2}= (116\,592\,030\pm150)\times 10^{-11}
\label{amu}
\end{equation}
lies slightly above theoretical evaluations based on the standard 
model, as much as $2.6\sigma$ in some cases.

It should be noted that all modern theoretical 
determinations~\cite{tres,four,six}
are compatible among themselves within errors (of order $100\times10^{-11}$) 
and that, with few exceptions, they are 
also compatible with the experimental 
result, at the  level of $1.5\sigma$ or less. Because of this, it is our feeling 
that a new, complete evaluation would be welcome since, 
in fact, there exists as yet no calculation  
that takes fully into account all theoretical constraints and
all the new experimental data. These  
data allow an improved evaluation of the low energy hadronic contributions 
to $a_\mu$, both directly from $e^+e^-$
annihilations (in the $\rho$ region~\cite{ocho} and around the $\omega$ and $\phi$ 
resonances~\cite{nine})  
and, indirectly, from $\tau$ decays~\cite{diez} and, also indirectly, 
from measurements of the pion form factor in the spacelike region~\cite{once}.
We have also used perturbative QCD in the region $s \geq 2\,\gev^2$
 (away from quark thresholds).
In particular, the recent LEP $\tau\to\nu_\tau+{\rm hadrons}$, 
and BES $e^+e^-\to{\rm hadrons}$ data justify the QCD result for 
$2\,\gev^2 \leq s\leq 3^2\,\gev^2$.
Moreover,  the BES~\cite{doce} data, covering 
$e^+e^-$ annihilations in the vicinity of $\bar{c}c$ threshold,  
permit a reliable evaluation of the corresponding hadronic pieces. 
Last but not least, the next order radiative corrections have 
been taken into account.

In the next sections we concentrate in the two most delicate pieces, from the point of
view of the final uncertainty: the pion form factor and the hadronic light-by-light
scattering contributions.


\subsection{The pion form factor}
To obtain $F_\pi(s)$ we fit the 
recent Novosibirsk data~\cite{ocho} on $e^+e^-\to\pi^+\pi^-$ and the 
tau decay data of Aleph and Opal~\cite{diez}. 
We take into account, at least partially, isospin breaking effects 
by allowing different masses and widths for the $\rho^0$, $\rho^+$ resonances. 
Moreover, and to get a good grip in the low energy region where
 data are inexistent or very poor, 
we also fit $F_\pi(s)$ at spacelike $s$~\cite{once}. 
This is possible in our approach because we use an expression for $F_\pi$
 that takes fully into account its analyticity properties.        
To be precise, we use  the Omn\`es-Muskhelishvili method. 
We write
$$F_\pi(s)=G(s)J(s).$$
Here $J$ is expressed in terms of the P-wave $\pi\pi$ phase shift, $\delta_1^1$, as 
$$J(s)=e^{1-\delta_1^1(s_0)/\pi}
\left(1-\frac{s}{s_0}\right)^{[1-\delta_1^1(s_0)/\pi]s_0/s}
\left(1-\frac{s}{s_0}\right)^{-1}
\exp\left\{\frac{s}{\pi}\int_{4m^2_\pi}^{s_0} \ ds'\;
\frac{\delta_1^1(s')}{s'(s'-s)}\right\}.$$
$s_0$ is the energy at which inelasticity starts becoming important (in practice, above the 
percent level); we will take $s_0=1.1\,\gev^2$ in actual calculations.

Because of the equality of the phase of $J(s)$ and the phase of $F_\pi(s)$ below $s=s_0$, it 
follows that $G(s)$ will be an analytic function also for 
$4m^2_\pi\leq s\leq s_0$, so in the whole $s$ plane except in a cut from 
$s=s_0$ to $+\infty$. If we now make the conformal transformation
$$z=\frac{\frac{1}{2}\sqrt{s_0}-\sqrt{s_0-s}}{\frac{1}{2}\sqrt{s_0}+\sqrt{s_0-s}}$$
then, as a function of $z$, $G$ will be analytic in the unit disk and we can thus 
write a convergent Taylor series for it. 
Incorporating the condition $G(0)=1$, that follows from $F_\pi(0)=1$, 
and undoing the transformation, we have
$$G(s)=1+
c_1\left[\frac{\frac{1}{2}\sqrt{s_0}-\sqrt{s_0-s}}{\frac{1}{2}\sqrt{s_0}+\sqrt{s_0-s}}
+\frac{1}{3}\right]+
c_2\Bigg[\left(\frac{\frac{1}{2}\sqrt{s_0}-\sqrt{s_0-s}}{\frac{1}{2}\sqrt{s_0}+\sqrt{s_0-s}}\right)^2
-\frac{1}{9}\Bigg]+\cdots,$$
$c_1,\,c_2,\,\dots$ free parameters. Actually, only two terms will be necessary to fit the data.

To obtain $J$,  
we use the effective range theory to write
$$\cot \delta_1^1(s)= \frac{s^{1/2}}{2k^3}(m^2_\rho-s)\hat{\psi}(s),\quad k=\frac{\sqrt{s-4m^2_\pi}}{2};$$
where we have extracted the zero corresponding to the rho resonance. 
The effective range function $\hat{\psi}(s)$ is analytic in the full $s$ plane
 except for a cut for $[-\infty,0]$ and the inelastic cut $[s_0,+\infty]$. 
We can profit from this analyticity by making again a conformal transformation
 into the unit circle, which is now given by
$$w=
\frac{\sqrt{s}-\sqrt{s_0-s}}{\sqrt{s}+\sqrt{s_0-s}}.$$
We  expand $\hat{\psi}$ in a convergent series of powers of $w$;
undoing the transformation we have
$$\delta_1^1(s)={\rm Arc\; cot}\left\{\frac{s^{1/2}}{2k^3}
(m^2_\rho-s)\left[b_0+b_1\frac{\sqrt{s}-\sqrt{s_0-s}}{\sqrt{s}+\sqrt{s_0-s}}+
\cdots\right]\right\}.$$
For the actual fits, only $b_0,\,b_1$, and $m_\rho$ are needed as 
parameters.

We give results both using only $F_\pi$ from 
$e^+e^-$ annihilations, or involving also the  
decay $\tau^+\to\bar{\nu}_\tau\pi^+\pi^0$, which last we consider to be our best estimates:

\begin{equation}
10^{11}\times a_{\mu}(2\pi;\,t\leq0.8\;\gev^2)=\;
\cases{
4\,774\pm31,\quad ({\rm TY1:}\ e^+e^-\,+\,\tau\,+\,{\rm spacelike}).\cr
4\,754\pm55,\quad ({\rm TY2:}\ e^+e^-\,+\,{\rm spacelike}).\cr
}
\label{rls}
\end{equation}

As a byproduct of the calculation we 
can also give precise values for the $\rho^0$, $\rho^+$ 
masses and widths or the pion electromagnetic radius; see~\cite{g2}.

\subsection{Hadronic light-by-light contributions}
A contribution to $a_{\mu}$ in a class by itself is the hadronic light-by-light scattering.
It can be evaluated  only using models. 
One can make a chiral model calculation, in the Nambu--Jona-Lasinio
version or the chiral perturbation theory variety, with a cut-off, or one can use a constituent 
quark model.
The result depends on the cut-off (for the chiral calculation) or on the constituent mass chosen 
for the quarks. What is worse, for the chiral model, 
a recent calculation~\cite{dseven} has challenged even its sign.
The variation from one model to the other is too large for comfort. 
We take as a typical value for the chiral calculation~\cite{tfour},
$$10^{11}\times a_{\mu}(\hbox{Hadronic light-by-light})=
-86\pm25\quad \hbox{(Chiral calculation; BPP, HKS)}.$$
In the calculation by~\cite{dseven} one simply reverses the sign.

For the constituent quark model we use the results
 of Laporta and Remiddi~\cite{tfive}. The contribution to $a_\mu$ of light-by-light scattering, 
with a loop with a fermion of  charge $Q_i$, and mass $m_i$ larger than the muon mass, is
$$10^{11}\times a_{\mu}(\hbox{Hadronic light-by-light})=
+92\pm20\quad \hbox{(Quark const. model)}$$
and the error is estimated by varying $m_{u,d}$ by 10\%. 
This is essentially the same result following the chiral model calculation 
by~\cite{dseven}.

Besides the sign problem, the
variations are unfortunate; one expects the chiral calculation to be valid for small 
values of the virtual photon momenta, 
and  the constituent model to hold for large values of the same.
Thus, almost half of the contribution to $a_{\mu}(\hbox{Hadronic light-by-light})$ 
in the chiral calculation comes from a region of momenta above $0.5$ GeV, 
where the chiral perturbation theory starts to fail,
while for this range of energies, and 
at least for the imaginary part of 
(diagonal) light-by-light scattering,  
the quark model reproduces reasonably well the experimental data.
Because of the uncertainties, we will give results with the extreme choices.

\subsection{Results}
Our results for the hadronic part of the anomaly 
depend on which model one believes for the 
hadronic light-by-light contribution, 
discussed in the previous section.
So we write,

\begin{equation}
10^{11}\times a_{\mu}(\hbox{Hadronic})=
\cases{
6\,993\pm69\quad({\rm TY1};\; {\rm Q.c.m.})\cr
6\,815\pm71\quad({\rm TY1};\; {\rm Ch.m. (BPP,HKS)})\cr
\phantom{x}\cr
6\,973\pm99\quad({\rm TY2};\; {\rm Q.c.m.})\cr
6\,795\pm100\quad({\rm TY2};\; {\rm Ch.m. (BPP,HKS)})\cr
}
\label{rsl}
\end{equation}

Note that 
in $a_{\mu}(\hbox{Hadronic})$ we include all 
hadronic contributions, $O(\alpha^3)$ as well as $O(\alpha^2)$.
In eqn.~(\ref{rsl}) ``Q.c.m." means that the light-by-light hadronic contribution was 
calculated with the quark constituent model, and ``Ch.m." that a chiral model was used. 
The errors include  the statistical errors, as well as the  
estimated systematic and theoretical ones. 
This is
to be compared with the value deduced from eqn.~(\ref{amu}) and electroweak corrections 
$$10^{11}\times a_{\mu}^{\rm exp.}(\hbox{Hadronic})=7\,174\pm150,$$
from which eqn.~(\ref{rsl}) differs by something between $1.1\sigma$ and $2.1\sigma$.
Our numbers are compared to other evaluations in Fig.~\ref{figura}.
Whether these results should be interpreted as providing 
further experimental validation of the standard model, or one can consider 
them as ``harbingers of new physics", we will 
leave for the reader to decide.

\FIGURE{
\mbox{
\epsfig{file=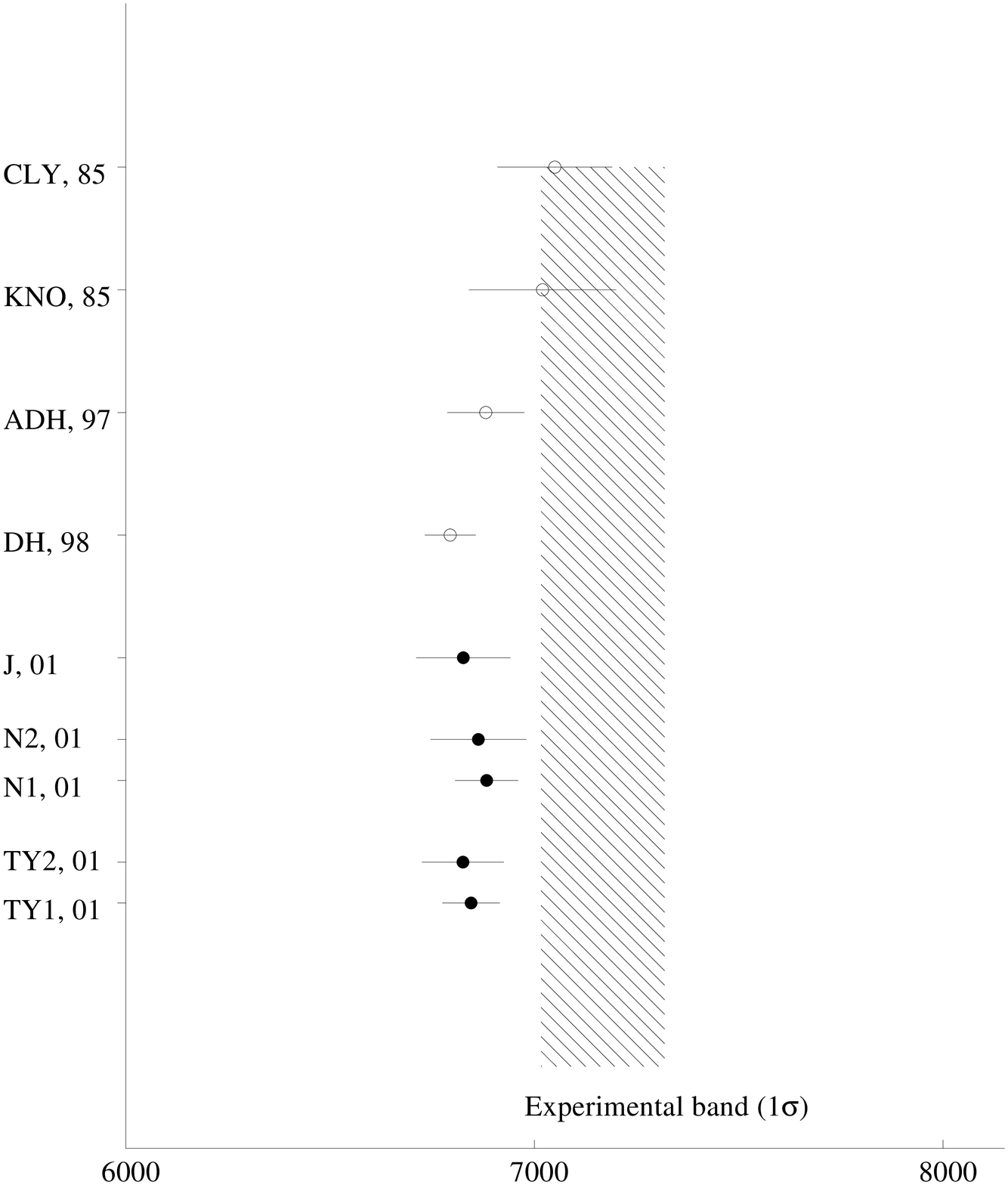,%
width=7.0cm,clip=}
\hskip 0.0cm
\epsfig{file=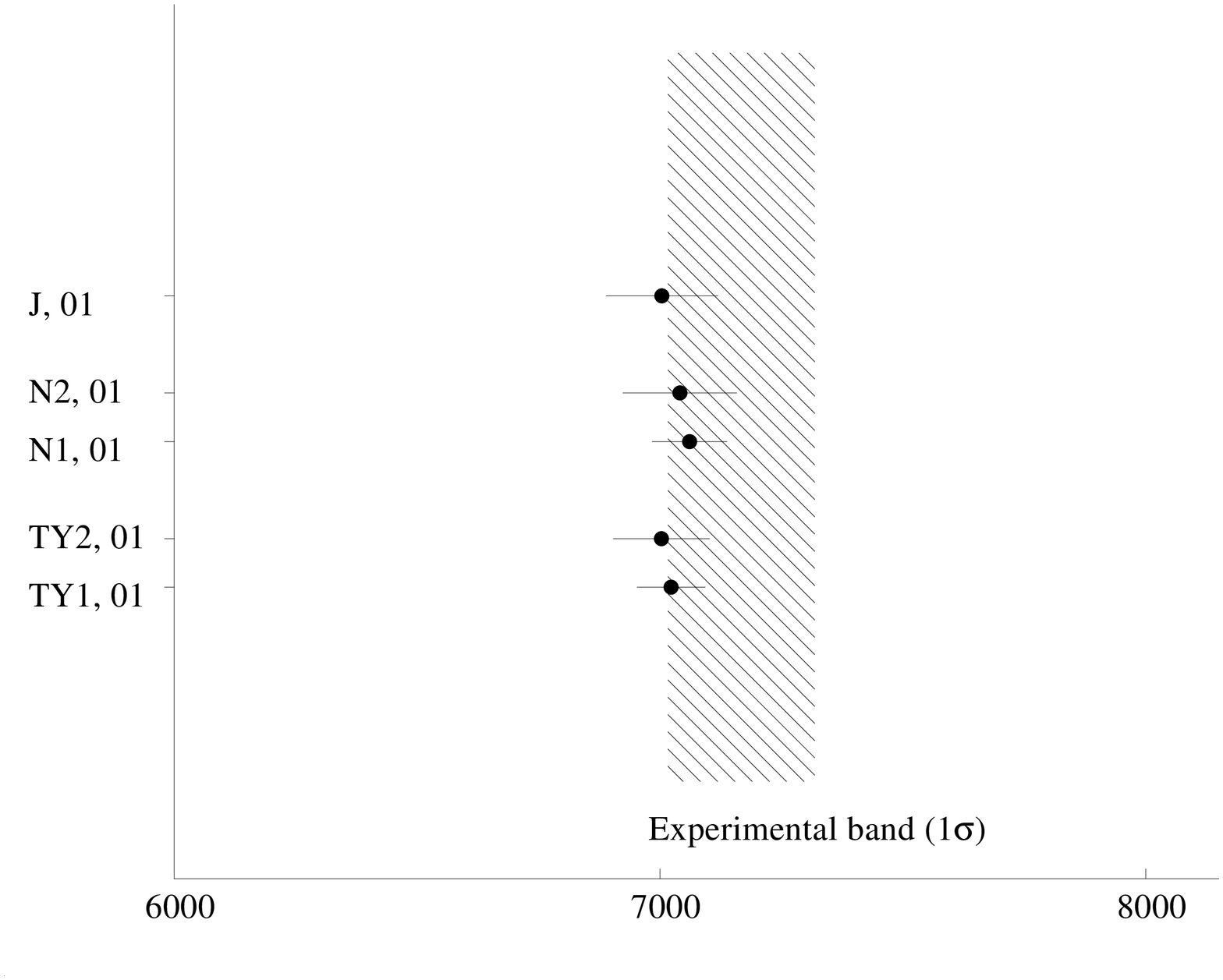,%
width=7.0cm,clip=}}
\caption{(Left) Theoretical results on $a_{\mu}({\rm Hadronic})\times10^{-11}$, and 
experiment. The chiral calculation result used for the hadronic light-by-light 
contribution. (Right) The same, but with the quark model result for the 
hadronic light-by-light contribution. Labels as in references~\cite{tres,four,six}.
TY1 and TY2 refer to the results in eqns.~(\ref{rls}) and (\ref{rsl}). \label{figura}}}

\section{Calculation of $\bar{\alpha}_{\rm Q.E.D.}$ on the $Z$}
With a simple change of integration kernel 
the previous analysis can be extended to evaluate the hadronic 
contribution to the QED running coupling, $\bar{\alpha}_{\rm Q.E.D.}(s)$, in 
particular for $s=M^2_Z$; an important quantity that enters 
into precision evaluations of electroweak observables.

By using a dispersion relation one can write this  hadronic 
contribution at energy squared $s$ as 
\begin{equation}
\Delta_{\rm had} \alpha(s) =-\frac{s\alpha}{3\pi}
\int_{4m^2_\pi}^\infty 
dt\;\frac{R(t)}{t(t-s)}
\label{bdr}
\end{equation}
with
$$R(t)=\frac{\sigma(e^+e^-\to{\rm hadrons};\,t)}{\sigma^{(0)}(e^+e^-\to\mu^+\mu^-;\,t)},
\quad \sigma^{(0)}(e^+e^-\to\mu^+\mu^-;\,t)\equiv\frac{4\pi\alpha^2}{3t}$$
and the integral in eqn.~(\ref{bdr}) has to be understood as a principal part integral.
This is similar to the Brodsky--de~Rafael expression for the hadronic 
contribution to the muon magnetic moment anomaly,
$$a_{\mu}(\hbox{h.v.p.})= \int_{4m^2_\pi}^\infty dt\,K(t) R(t).$$

Therefore, we can carry over all the work from the previous section 
with the simple replacement   
$$K(t)\to-\frac{s\alpha}{3\pi}\frac{1}{t(t-s)}.$$
We find to next to leading order in $\alpha$,
$$10^5\times\Delta_{\rm had}\alpha(M_Z^2)=
2740\pm12$$
or, excluding the top quark contribution,
$$10^5\times\Delta_{\rm had}\alpha^{(5)}(M_Z^2)=
2747\pm12.$$

Other recent determinations can be found in~\cite{six,dseis}.

\section{Conclusions}
In conclusion, we have performed a detailed evaluation of the hadronic contributions to the muon 
$g_{\mu}-2$~\cite{g2}, and to the running electromagnetic coupling~\cite{az}.

In the case of the muon $g_{\mu}-2$ an important limitation of the attainable accuracy is the 
light-by-light scattering contribution; in particular, a third chiral model 
calculation is required to clear the sign controversy.

\end{document}